\RequirePackage{ifpdf}
\ifpdf % We are running pdfTeX in pdf mode
\documentclass[pdftex]{sigma}
\else
\documentclass{sigma}
\fi

\begin{document}
\allowdisplaybreaks

\renewcommand{\PaperNumber}{009}

\FirstPageHeading

\ShortArticleName{On Action Invariance under Linear Spinor-Vector
Supersymmetry}

\ArticleName{On Action Invariance under Linear Spinor-Vector\\
Supersymmetry}

\Author{Kazunari SHIMA and Motomu TSUDA}
\AuthorNameForHeading{K.~Shima and M.~Tsuda} \Address{Laboratory
of Physics, Saitama Institute of Technology, Okabe-machi, Saitama
369-0293, Japan}
\Email{\href{mailto:shima@sit.ac.jp}{shima@sit.ac.jp},
\href{mailto:tsuda@sit.ac.jp}{tsuda@sit.ac.jp}}

\ArticleDates{Received October 21, 2005, in final form January 10,
2006; Published online January 24, 2006}

\Abstract{We show explicitly that a free Lagrangian expressed in
terms of scalar, spinor, vector and Rarita--Schwinger (RS) fields
is invariant under linear supersymmetry transformations generated
by a global spinor-vector parameter. A (generalized) gauge
invariance of the Lagrangian for the RS field is also discussed.}

\Keywords{spinor-vector supersymmetry; Rarita--Schwinger field}

\Classification{81T60}

Both linear (L) \cite{WZ} and nonlinear (NL) \cite{VA}
supersymmetry (SUSY) are realized based on a SUSY algebra where
spinor generators are introduced in addition to Poincar\'e
generators. The relation between the L and the NL SUSY, i.e., the
algebraic equivalence between various (renormalizable)
spontaneously broken L supermultiplets and a NL SUSY action
\cite{VA} in terms of a Nambu-Goldstone (NG) fermion has been
investigated by many authors \cite{IK,Ro,UZ,STT1}.

An extension of the Volkov--Akulov (VA) model~\cite{VA} of NL SUSY
based on a spinor-vector generator, called the spin-3/2 SUSY,
hitherto, and its NL realization in terms of a spin-3/2 NG fermion
have been constructed by N.S.~Baaklini~\cite{Ba}. From the
spin-3/2 NL SUSY model, L realizations of the spin-3/2 SUSY are
suggested as corresponding supermultiplets to a spin-3/2 NL SUSY
action~\cite{Ba} through a linearization. The linearization of the
spin-3/2 NL SUSY is also useful from the viewpoint towards
constructing a SUSY composite unified theory based on $SO(10)$
super-Poincar\'e (SP) group (the superon-graviton model (SGM))
\cite{KS,ST_32SGM}, and it may give new insight into an analogous
mechanism with the super-Higgs one~\cite{DZ} for high spin fields
which appear in SGM (up to spin-3 fields).

Recently, we have studied the unitary representation of the
spin-3/2 SUSY algebra in \cite{Ba} towards the linearization of
the spin-3/2 NL SUSY \cite{ST}. Since the spinor-vector generator
has the role of creation and annihilation operators which raise or
lower the helicity of states by 1/2 or by 3/2, the structure of
the (physical) L supermultiplets induced from the spin-3/2 SUSY
algebra is shown for example as
\begin{gather}
\left[ \ \underline{1} \left( +{3 \over 2} \right), \underline{2}
(+1), \underline{1} \left( +{1 \over 2} \right), \underline{1}
(0), \underline{2} \left( -{1 \over 2} \right), \underline{1} (-1)
\ \right] + [\ {\rm CPT\ conjugate}\ ] \label{irrep}
\end{gather}
for the massless case. In equation~(\ref{irrep}) $\underline{n}
(\lambda)$ means the number of states $n$ for the helicity
$\lambda$. Therefore, it is expected in the above examples that
the spin-3/2 L supermultiplets contain scalar, spinor, vector and
Rarita--Schwinger (RS) fields as fundamental fields. In order to
explicitly show that those fields constitute the spin-3/2 L
supermultiplet, we have to prove an action invariance under
appropriate spin-3/2 L SUSY transformations whose commutator
algebras close as a representation of the Baaklini's spin-3/2 SUSY
algebra. Namely, we have to determine the form of the spin-3/2 L
SUSY transformations both from the action invariance and from the
closure of those commutator algebra.

In this paper, as a first step to do these calculations we
explicitly demonstrate the spin-3/2 L SUSY invariance of a free
Lagrangian in terms of spin-$(0^\pm, 1/2, 1, 3/2)$ fields, and we
discuss the spin-3/2 L SUSY transformations determined from the
invariance of the Lagrangian. Here we just mention the relation to
the so-called no-go theorem~\cite{CM,HLS} based upon the
$S$-matrix arguments, i.e. the case for the S-matrix (the true
vacuum) is well defined. (Note that the vacuum of NLSUSY VA model
may have rich structures, for $N=1$ VA model is equivalent to
$N=1$ LSUSY scalar supermultiplet and also to $N=1$ LSUSY axial
vector supermultiplet as we have proved.) We discuss in this paper
the {\it global} L SUSY with spin-3/2 charges for the {\it free}
Lagrangian, which are free from the no-go theorem, so far.
Those are important preliminaries not only to find out a
(spontaneously broken) LSUSY supermultiplet, which is equivalent
to the NL realization of the spin-3/2 SUSY algebra~\cite{Ba}, but
also to obtain some information for linearizing the {\it
interacting global} NL SUSY theory with spin-3/2 (NG) fields in
curved spacetime (i.e., the spin-3/2 SGM)~\cite{ST_32SGM}. From
these viewpoints we think it is worthwhile presenting the progress
report along this direction.

Let us denote spin-$(0^\pm, 1/2, 1, 3/2)$ fields beside auxiliary
fields as follows: namely, $A$ and $B$ for scalar fields,
$\lambda$ for a (Majorana) spinor, $v_a$ for a $U(1)$ gauge field
and $\lambda_a$ for a (Majorana) RS field. For these component
fields we consider a parity conserving free Lagrangian given
by\footnote{Minkowski spacetime indices are denoted by $a, b,
\ldots = 0, 1, 2, 3$. The Minkowski spacetime metric is ${1 \over
2}\{ \gamma^a, \gamma^b \} = \eta^{ab} = (+, -, -, -)$ and
$\sigma^{ab} = {i \over 4}[\gamma^a, \gamma^b]$.}
\begin{gather}
L  = {1 \over 2} (\partial_a A)^2 + {1 \over 2} (\partial_a B)^2 +
{i \over 2} \bar\lambda \!\!\not\!\partial \lambda - {1 \over 4}
(F_{ab})^2 + {i \over 2} X_1 \bar\lambda^a \!\!\not\!\partial
\lambda_a + {i \over 2} X_2 (\bar\lambda^a \gamma^b \partial_a
\lambda_b + \bar\lambda^a \gamma_a \partial^b \lambda_b)
\nonumber \\
\phantom{L  =}{}- {1 \over 2} X_3 \epsilon_{abcd} \bar\lambda^a
\gamma_5 \gamma^b \partial^c \lambda^d + Y_1 \bar\lambda
\partial^a \lambda_a + i Y_2 \bar\lambda \sigma^{ab} \partial_a
\lambda_b, \label{32LSUSYaction}
\end{gather}
where $F_{ab} = \partial_a v_b - \partial_b v_a$, and $X_i$ (for
$i = 1, 2, 3$) with $X_3 = 1 - X_1$ and $Y_i$ (for $i = 1, 2$) are
arbitrary real parameters. Note that in
equation~(\ref{32LSUSYaction}) the general form of the Lagrangian
for the RS field $\lambda_a$ is adopted, and also the derivative
coupling kinetic-like terms expressed in terms of $\lambda$ and
$\lambda_a$, as the last two terms are introduced without the loss
of generality.

Furthermore, we define spin-3/2 L SUSY transformations generated
by a global (Majorana) spinor-vector parameter $\zeta_a$ as
\begin{gather}
\delta_Q A  = i \alpha  \bar\zeta^a \gamma_a \lambda + a_1
\bar\zeta^a \lambda_a + i a_2  \bar\zeta_a \sigma^{ab} \lambda_b,
\label{32LSUSYtransfn-0}\\
\delta_Q B  = \alpha'  \bar\zeta^a \gamma_5 \gamma_a \lambda + i
a'_1  \bar\zeta^a \gamma_5 \lambda_a
+ a'_2  \bar\zeta_a \gamma_5 \sigma^{ab} \lambda_b, \\
\delta_Q v_a  = \alpha''_1  \bar\zeta_a \lambda + i \alpha''_2
\bar\zeta^b \sigma_{ab} \lambda + i a''_1  \bar\zeta_a \gamma^b
\lambda_b + i a''_2  \bar\zeta^b \gamma_a \lambda_b + i a''_3
\bar\zeta^b \gamma_b \lambda_a
+ a''_4  \epsilon_{abcd} \bar\zeta^b \gamma_5 \gamma^c \lambda^d ,\\
\delta_Q \lambda  = \beta_1  \zeta^a \partial_a A + i \beta_2
\sigma^{ab} \zeta_a \partial_b A + i \beta'_1  \gamma_5 \zeta^a
\partial_a B + \beta'_2  \gamma_5 \sigma^{ab} \zeta_a \partial_b B
\nonumber \\
\phantom{\delta_Q \lambda  =}{} + i \beta''_1  \gamma^a \zeta^b
F_{ab} + {1 \over 2} \beta''_2  \epsilon_{abcd} \gamma_5 \gamma^a
\zeta^b F^{cd}, \label{32LSUSYtransfn-12}
\\
\delta_Q \lambda_a  =
 i b_1  \gamma_a \zeta^b \partial_b A
+ i b_2  \gamma^b \zeta_a \partial_b A + i b_3  \gamma^b \zeta_b
\partial_a A + b_4  \epsilon_{abcd} \gamma_5 \gamma^b \zeta^c
\partial^d A
\nonumber \\
\phantom{\delta_Q \lambda_a  =}{} + b'_1  \gamma_5 \gamma_a
\zeta^b \partial_b B + b'_2  \gamma_5 \gamma^b \zeta_a \partial_b
B + b'_3  \gamma_5 \gamma^b \zeta_b \partial_a B + i b'_4
\epsilon_{abcd} \gamma^b \zeta^c \partial^d B
\nonumber \\
\phantom{\delta_Q \lambda_a  =}{} + b''_1  \zeta^b F_{ab} + i
b''_2  \sigma_a{}^b \zeta^c F_{bc} + {i \over 2} b''_3
\sigma^{bc} \zeta_a F_{bc} + i b''_4  \sigma^{bc} \zeta_b F_{ac} +
{i \over 2} b''_5  \epsilon_{abcd} \gamma_5 \zeta^b F^{cd},
\label{32LSUSYtransfn-32}
\end{gather}
where the $\alpha$, $\alpha'$, $\alpha''_i$ (for $i = 1, 2$),
$\beta_i$ (for $i = 1, 2$), $\beta'_i$ (for $i = 1, 2$),
$\beta''_i$ (for $i = 1, 2$), $a_i$ (for $i = 1, 2$), $a'_i$ (for
$i = 1, 2$), $a''_i$ (for $i = 1, \ldots, 4$), $b_i$ (for $i = 1,
\ldots, 4$), $b'_i$ (for $i = 1, \ldots, 4$) and $b''_i$ (for $i =
1, \ldots, 5$) are also arbitrary real parameters. The values of
those parameters in equation~(\ref{32LSUSYaction}) and in
equations from (\ref{32LSUSYtransfn-0}) to
(\ref{32LSUSYtransfn-32}) are determined from conditions for the
spin-3/2 SUSY invariance of the Lagrangian (\ref{32LSUSYaction})
as is shown below.

Application of the spin-3/2 SUSY transformations to equation
(\ref{32LSUSYaction}) (\ref{32LSUSYtransfn-0}) to
(\ref{32LSUSYtransfn-32}) gives various terms as
\begin{gather}
\delta_Q L =
        F_1 (\bar\zeta^a \lambda_a \Box A,
        \bar\zeta^a \lambda^b \partial_a \partial_b A,
        \bar\zeta^a \sigma_{ab} \lambda^b \Box A,
        \bar\zeta^a \sigma^{bc} \lambda_c \partial_a \partial_b A,
        \bar\zeta^a \sigma_{ab} \lambda_c \partial^b \partial^c A)
\nonumber \\
\phantom{\delta_Q L =}{}
         +  F_2 (\bar\zeta^a \gamma_5 \lambda_a \Box B,
        \bar\zeta^a \gamma_5 \lambda^b \partial_a \partial_b B,
        \bar\zeta^a \gamma_5 \sigma_{ab} \lambda^b \Box B,
      \bar\zeta^a \gamma_5 \sigma^{bc} \lambda_c \partial_a \partial_b B,
      \bar\zeta^a \gamma_5 \sigma_{ab} \lambda_c \partial^b \partial^c B)
\nonumber \\
\phantom{\delta_Q L =}{} +  F_3 (\bar\zeta^a \gamma_a \lambda \Box
A,
      \bar\zeta^a \gamma^b \lambda \partial_a \partial_b A)
 +  F_4 (\bar\zeta^a \gamma_5 \gamma_a \lambda \Box B,
      \bar\zeta^a \gamma_5 \gamma^b \lambda \partial_a \partial_b B)
\nonumber \\
\phantom{\delta_Q L =}{} +  F_5 (\bar\zeta^a \gamma^b \lambda_b
\partial^c F_{ac},
      \bar\zeta^a \gamma^b \lambda_a \partial^c F_{bc},
      \bar\zeta^a \gamma_a \lambda^b \partial^c F_{bc},
      \bar\zeta^a \gamma^b \lambda^c \partial_c F_{ab},
      \bar\zeta^a \gamma^b \lambda^c \partial_a F_{bc},
\nonumber \\
\phantom{\delta_Q L =+  F_5 (}{} \epsilon^{abcd} \bar\zeta^e
\gamma_5 \gamma_a \lambda_b \partial_e F_{cd}, \epsilon^{abcd}
\bar\zeta_a \gamma_5 \gamma^e \lambda_b \partial_e F_{cd},
\epsilon^{abcd} \bar\zeta_a \gamma_5 \gamma_b \lambda^e \partial_e
F_{cd})
\nonumber \\
\phantom{\delta_Q L =}{} +  F_6 (\bar\zeta^a \lambda \partial^b
F_{ab},
      \bar\zeta^a \sigma_{ab} \lambda \partial_c F^{bc},
      \bar\zeta^a \sigma^{bc} \lambda \partial_b F_{ac})
 + \ [\ {\rm tot.\ der.\ terms}\ ],
\label{variation}
\end{gather}
where we have used the relation, $\partial_a F_{bc} + \partial_c
F_{ab} + \partial_b F_{ca} = 0$. Therefore, the conditions for
$\delta_Q L = 0$ (up to total derivative terms) are as follows;
namely, the vanishing conditions of coefficients for the each kind
of the terms in equation~(\ref{variation}) are
\begin{gather}
a_1 + X_1 b_2 + X_2 b_3 + 2 X_3 b_4 - {1 \over 4} Y_2 \beta_2 = 0,
\nonumber \\
(X_1 + 5 X_2) b_1 + 2 X_2 b_2 + (X_1 + X_2) b_3 - 2 X_3 b_4 + Y_1
\beta_1 + {1 \over 4} Y_2 \beta_2 = 0,
\nonumber \\
a_2 + 2 X_3 b_2 - 2 X_2 b_3 + 2 (X_1 - X_3) b_4 + {1 \over 2} Y_2
\beta_2 = 0,
\nonumber \\
(X_1 - X_2 - 2 X_3) b_1 - (X_2 + X_3) b_2 - (X_1 - X_3) b_4 + {1
\over 2} Y_2 \left( \beta_1 - {1 \over 2} \beta_2 \right) = 0,
\nonumber \\
(X_2 - X_3) b_2 - (X_1 + X_2) b_3 - (X_1 + 2 X_2 - X_3) b_4 - {1
\over 2} \left( Y_1 + {1 \over 2} Y_2 \right) \beta_2 = 0
\label{conditions-F1}
\end{gather}
for the terms in $F_1$,
\begin{gather}
a'_1 + X_1 b'_2 + X_2 b'_3 - 2 X_3 b'_4 + {1 \over 4} Y_2 \beta'_2
= 0,
\nonumber \\
(X_1 + 5 X_2) b'_1 + 2 X_2 b'_2 + (X_1 + X_2) b'_3 + 2 X_3 b'_4 +
Y_1 \beta'_1 - {1 \over 4} Y_2 \beta'_2 = 0,
\nonumber \\
- a'_2 + 2 X_3 b'_2 - 2 X_2 b'_3 - 2 (X_1 - X_3) b'_4 - {1 \over
2} Y_2 \beta'_2 = 0,
\nonumber \\
(X_1 - X_2 - 2 X_3) b'_1 - (X_2 + X_3) b'_2 + (X_1 - X_3) b'_4 +
{1 \over 2} Y_2 \left( \beta'_1 + {1 \over 2} \beta'_2 \right) =
0,
\nonumber \\
(X_2 - X_3) b'_2 - (X_1 + X_2) b'_3 + (X_1 + 2 X_2 - X_3) b'_4 +
{1 \over 2} \left( Y_1 + {1 \over 2} Y_2 \right) \beta'_2 = 0
\label{conditions-F2}
\end{gather}
for the terms in $F_2$,
\begin{gather}
2 \alpha + \beta_2 + Y_2 b_2 + 2 Y_1 b_3 - 2 Y_2 b_4 = 0,
\nonumber \\
2 \beta_1 - \beta_2 + (2 Y_1 - 3 Y_2) b_1 + (2 Y_1 - Y_2) b_2 + 2
Y_2 b_4 = 0 \label{conditions-F3}
\end{gather}
for the terms in $F_3$,
\begin{gather}
- 2 \alpha' - \beta'_2 + Y_2 b'_2 + 2 Y_1 b'_3 + 2 Y_2 b'_4 = 0,
\nonumber \\
2 \beta'_1 + \beta'_2 + (2 Y_1 - 3 Y_2) b'_1 + (2 Y_1 - Y_2) b'_2
- 2 Y_2 b'_4 = 0 \label{conditions-F4}
\end{gather}
for the terms in $F_4$,
\begin{gather}
2 a''_1 - 2 X_2 b''_1 - (X_1 - 2 X_3) b''_2 - X_3 b''_3 - 2 X_3
b''_4 - Y_2 \beta''_1 = 0,
\nonumber \\
2 a''_2 + X_1 b''_3 + (X_2 + X_3) b''_4 + 2 X_3 b''_5 + Y_2
\beta''_2 = 0,
\nonumber \\
2 a''_3 + X_3 b''_3 + (X_1 - X_2) b''_4 - 2 X_3 b''_5 - Y_2
\beta''_2 = 0,
\nonumber \\
2 (X_1 + X_2) b''_1 - (X_1 + 3 X_2 - 2 X_3) b''_2 + (X_2 - X_3)
b''_3 + (X_2 - X_3) b''_4 - (2 Y_1 + Y_2) \beta''_1 = 0,
\nonumber \\
2 X_1 b''_1 + (X_2 + 2 X_3) b''_2 - (X_2 + X_3) b''_3 - X_1 b''_4
+ 2 X_3 b''_5 - Y_2 \beta''_1 + Y_2 \beta''_2 = 0,
\nonumber \\
2 a''_4 + 2 X_3 b''_1 - (X_1 - X_2 - X_3) b''_2 + X_2 b''_3 + X_2
b''_4 + Y_2 \beta''_1 = 0,
\nonumber \\
2 a''_4 - X_3 b''_3 + (X_2 + X_3) b''_4 - 2 X_1 b''_5 + Y_2
\beta''_2 = 0,
\nonumber \\
2 a''_4 - X_2 b''_3 + (X_1 + 3 X_2) b''_4 + 2 X_2 b''_5 - 2 Y_1
\beta''_2 = 0
\end{gather}
for the terms in $F_5$, and
\begin{gather}
4 (\alpha''_1 - \beta''_1) + 4 Y_1 b''_1 + 3 Y_2 b''_2 - Y_2 b''_3
- Y_2 b''_4 = 0,
\nonumber \\
2 (\alpha''_2 - 2 \beta''_2) + Y_2 b''_3 + (2 Y_1 + Y_2) b''_4 - 2
Y_2 b''_5 = 0,
\nonumber \\
4 (\beta''_1 - \beta''_2) + 2 Y_2 b''_1 + 2 (Y_1 - Y_2) b''_2 - (2
Y_1 - Y_2) b''_3 - Y_2 b''_4 - 2 Y_2 b''_5 = 0
\label{conditions-F6}
\end{gather}
for the terms in $F_6$. Up to the above arguments we can easily
observe the existence of the nontrivial solutions for equations
from (\ref{conditions-F1}) to (\ref{conditions-F6}).

Note that if we choose tentatively the arbitrary parameters as
\begin{gather}
a'_1 = a_1, \qquad a'_2 = - a_2,  \qquad b'_i = b_i \quad ({\rm
for} \ i = 1, 2, 3), \qquad b'_4 = - b_4,
\nonumber \\
\alpha' = - \alpha, \qquad \beta'_1 = \beta_1, \qquad \beta'_2 = -
\beta_2, \label{choose}
\end{gather}
then the conditions in equation~(\ref{conditions-F2}) and
(\ref{conditions-F4}) are equal to those in
equation~(\ref{conditions-F1}) and (\ref{conditions-F3}),
respectively. We can find solutions of the parameters which
satisfy the conditions (\ref{conditions-F1}) to
(\ref{conditions-F6}) with equation~(\ref{choose}) , i.e., it can
be shown that the Lagrangian (\ref{32LSUSYaction}) is invariant
under the spin-3/2 SUSY transformation (\ref{32LSUSYtransfn-0}) to
(\ref{32LSUSYtransfn-32}).

Here we notice a special example of solutions for the conditions
(\ref{conditions-F1}) to (\ref{conditions-F6}), which is given by
$X_1 = X_2 = Y_1 = Y_2 = 0$ (it automatically gives $X_3 = 0$ as
is understood from the second equation in
equation~(\ref{conditions-F1})). This example means that the RS
field does not contribute to equation~(\ref{32LSUSYaction}), and
then the free Lagrangian for the spin-$(0^\pm, 1/2, 1)$ fields is
spin-3/2 SUSY invariant under $\beta_2 = 2 \beta_1 = - 2 \alpha$
($\beta'_2 = - 2 \beta'_1 = - 2 \alpha'$) and $\beta''_2 =
\beta''_1 = (1/2) \alpha''_2 = \alpha''_1$. (However, in this case
commutator algebras for the spin-3/2 SUSY transformations
(\ref{32LSUSYtransfn-0}) to (\ref{32LSUSYtransfn-12}) do not close
as a~spin-3/2 SUSY representation of the Baaklini's
type~\cite{Ba}.)

Let us also discuss on the invariance of the Lagrangian
(\ref{32LSUSYaction}) under the gauge transformation of the RS
field. We define the (generalized) gauge transformation of
$\lambda_a$ generated by a local spinor parameter $\epsilon$ as
\begin{gather}
\delta_g \lambda_a = p  \partial_a \epsilon + i q  \sigma_{ab}
\partial^b \epsilon, \label{gauge}
\end{gather}
where $p$ and $q$ are arbitrary (real) parameters. The variation
of equation~(\ref{32LSUSYaction}) with respect to
equation~(\ref{gauge}) becomes
\begin{gather}
\delta_g L
 =  i \left\{ (X_1 + X_2) p - {1 \over 2} (X_1 + 3 X_2 - 2 X_3) q \right\}
\bar\lambda^a \!\!\not\!\partial \partial_a \epsilon\! + i \left\{
X_2 p + {1 \over 2} (X_1 - 2 X_3) q \right\} \bar\lambda^a
\gamma_a \Box \epsilon \!
\nonumber \\
\phantom{\delta_g L  = }{} + \left( Y_1 p + {3 \over 4} Y_2 q
\right) \bar\lambda \Box \epsilon + [\ {\rm tot.\ der.\ terms}\ ].
\label{variation-gauge}
\end{gather}

From equation~(\ref{variation-gauge}) the conditions for $\delta_g
L = 0$ (up to total derivative terms) are read as
\begin{gather}
X_1^2 + 3 X_2^2 + 2 X_1 X_2 - 4 X_2 X_3 - 2 X_1 X_3 = 0,
\nonumber \\
4 Y_1 \ p + 3 Y_2 \ q = 0. \label{conditions-gauge}
\end{gather}
Therefore, the Lagrangian for $\lambda_a$ in
equation~(\ref{32LSUSYaction}) is invariant under
equation~(\ref{gauge}) for arbitrary values of $X_i$, $Y_i$, $p$
and $q$ which satisfy equation~(\ref{conditions-gauge}).

It can be also shown that the Lagrangian (\ref{32LSUSYaction}) is
invariant under both the spin-3/2 SUSY transformations
(\ref{32LSUSYtransfn-0}) to (\ref{32LSUSYtransfn-32}) and the
gauge transformation (\ref{gauge}). We need further investigations
on the closure of commutator algebras for equations from
(\ref{32LSUSYtransfn-0}) to (\ref{32LSUSYtransfn-32}) as a
representation of the spin-3/2 SUSY algebra in~\cite{Ba}.

\LastPageEnding

\end{document}